
==================




 \font\twelverm=cmr10 scaled 1200    \font\twelvei=cmmi10 scaled 1200
 \font\twelvesy=cmsy10 scaled 1200   \font\twelveex=cmex10 scaled 1200
 \font\twelvebf=cmbx10 scaled 1200   \font\twelvesl=cmsl10 scaled 1200
 \font\twelvett=cmtt10 scaled 1200   \font\twelveit=cmti10 scaled 1200

 \skewchar\twelvei='177   \skewchar\twelvesy='60


 \def\twelvepoint{\normalbaselineskip=12.4pt
   \abovedisplayskip 12.4pt plus 3pt minus 9pt
   \belowdisplayskip 12.4pt plus 3pt minus 9pt
   \abovedisplayshortskip 0pt plus 3pt
   \belowdisplayshortskip 7.2pt plus 3pt minus 4pt
   \smallskipamount=3.6pt plus1.2pt minus1.2pt
   \medskipamount=7.2pt plus2.4pt minus2.4pt
   \bigskipamount=14.4pt plus4.8pt minus4.8pt
   \def\rm{\fam0\twelverm}          \def\it{\fam\itfam\twelveit}%
   \def\sl{\fam\slfam\twelvesl}     \def\bf{\fam\bffam\twelvebf}%
   \def\mit{\fam 1}                 \def\cal{\fam 2}%
   \def\tt{\twelvett}
   \def\nullspace{\nulldelimiterspace=0pt \mathsurround=0pt }
   \def\big##1{{\hbox{$\left##1\vbox to 10.2pt{}\right.\nullspace$}}}
   \def\Big##1{{\hbox{$\left##1\vbox to 13.8pt{}\right.\nullspace$}}}
   \def\bigg##1{{\hbox{$\left##1\vbox to 17.4pt{}\right.\nullspace$}}}
   \def\Bigg##1{{\hbox{$\left##1\vbox to 21.0pt{}\right.\nullspace$}}}
   \textfont0=\twelverm   \scriptfont0=\tenrm   \scriptscriptfont0=\sevenrm
   \textfont1=\twelvei    \scriptfont1=\teni    \scriptscriptfont1=\seveni
   \textfont2=\twelvesy   \scriptfont2=\tensy   \scriptscriptfont2=\sevensy
   \textfont3=\twelveex   \scriptfont3=\twelveex  \scriptscriptfont3=\twelveex
   \textfont\itfam=\twelveit
   \textfont\slfam=\twelvesl
   \textfont\bffam=\twelvebf \scriptfont\bffam=\tenbf
   \scriptscriptfont\bffam=\sevenbf
   \normalbaselines\rm}


 \def\tenpoint{\normalbaselineskip=12pt
   \abovedisplayskip 12pt plus 3pt minus 9pt
   \belowdisplayskip 12pt plus 3pt minus 9pt
   \abovedisplayshortskip 0pt plus 3pt
   \belowdisplayshortskip 7pt plus 3pt minus 4pt
   \smallskipamount=3pt plus1pt minus1pt
   \medskipamount=6pt plus2pt minus2pt
   \bigskipamount=12pt plus4pt minus4pt
   \def\rm{\fam0\tenrm}          \def\it{\fam\itfam\tenit}%
   \def\sl{\fam\slfam\tensl}     \def\bf{\fam\bffam\tenbf}%
   \def\smc{\tensmc}             \def\mit{\fam 1}%
   \def\cal{\fam 2}%
   \textfont0=\tenrm   \scriptfont0=\sevenrm   \scriptscriptfont0=\fiverm
   \textfont1=\teni    \scriptfont1=\seveni    \scriptscriptfont1=\fivei
   \textfont2=\tensy   \scriptfont2=\sevensy   \scriptscriptfont2=\fivesy
   \textfont3=\tenex   \scriptfont3=\tenex     \scriptscriptfont3=\tenex
   \textfont\itfam=\tenit
   \textfont\slfam=\tensl
   \textfont\bffam=\tenbf \scriptfont\bffam=\sevenbf
   \scriptscriptfont\bffam=\fivebf
   \normalbaselines\rm}


 \def\beginlinemode{\endmode
   \begingroup\parskip=0pt \obeylines\def\\{\par}\def\endmode{\par\endgroup}}
 \def\beginparmode{\endmode
   \begingroup \def\endmode{\par\endgroup}}
 \let\endmode=\par
 {\obeylines\gdef\
 {}}
 \def\singlespace{\baselineskip=\normalbaselineskip}
 
 \def\oneandahalfspace{\baselineskip=\normalbaselineskip
   \multiply\baselineskip by 3 \divide\baselineskip by 2}
 \def\doublespace{\baselineskip=\normalbaselineskip \multiply\baselineskip by
2}
 
 \newcount\firstpageno
 \firstpageno=2

\footline={\ifnum\pageno<\firstpageno{\hfil}\else{\hfil\twelverm\folio\hfil}\fi}

 \let\rawfootnote=\footnote              
 \def\footnote#1#2{{\rm\singlespace\parindent=0pt\rawfootnote{#1}{#2}}}
 \def\raggedcenter{\leftskip=4em plus 12em \rightskip=\leftskip
   \parindent=0pt \parfillskip=0pt \spaceskip=.3333em \xspaceskip=.5em
   \pretolerance=9999 \tolerance=9999
   \hyphenpenalty=9999 \exhyphenpenalty=9999 }
 \def\dateline{\rightline{\ifcase\month\or
   January\or February\or March\or April\or May\or June\or
   July\or August\or September\or October\or November\or December\fi
   \space\number\year}}
 \def\received{\vskip 3pt plus 0.2fill
 \centerline{\sl (Received\space\ifcase\month\or
   January\or February\or March\or April\or May\or June\or
   July\or August\or September\or October\or November\or December\fi
   \qquad, \number\year)}}


 \hsize=6.5truein
 \vsize=8.75truein
 \parskip=\medskipamount
 \twelvepoint            
 \doublespace            
 \overfullrule=0pt       



 \def\title                      
   {\null\vskip 3pt plus 0.2fill
    \beginlinemode \doublespace \raggedcenter \bf}

 \def\author                     
   {\vskip 3pt plus 0.2fill \beginlinemode
    \singlespace \raggedcenter}

 \def\affil                      
   {\vskip 3pt plus 0.1fill \beginlinemode
    \oneandahalfspace \raggedcenter \sl}

 \def\abstract                   
   {\vskip 3pt plus 0.3fill \beginparmode
    \doublespace \narrower ABSTRACT: }

 \def\endtitlepage               
   {\endpage                     
    \body}

 \def\body                       
   {\beginparmode}               

 \def\head#1{                    
   \filbreak\vskip 0.5truein     
   {\immediate\write16{#1}
    \raggedcenter \uppercase{#1}\par}
    \nobreak\vskip 0.25truein\nobreak}

 \def\refto#1{$^{#1}$}           

 \def\references                 
   {\head{References}            

    \beginparmode
    \frenchspacing \parindent=0pt \leftskip=1truecm
    \parskip=8pt plus 3pt \everypar{\hangindent=\parindent}}

 \gdef\refis#1{\indent\hbox to 0pt{\hss#1.~}}    

 \gdef\journal#1, #2, #3, 1#4#5#6{               

     {\sl #1~}{\bf #2}, #3, (1#4#5#6)}           

 \gdef\journ2 #1, #2, #3, 1#4#5#6{               

     {\sl #1~}{\bf #2}: #3, (1#4#5#6)}           

 \def\refstylenp{                
   \gdef\refto##1{ (##1)}                                
   \gdef\refis##1{\indent\hbox to 0pt{\hss##1)~}}        
   \gdef\journal##1, ##2, ##3, ##4 {                     
      {\sl ##1~}{\bf ##2~}(##3) ##4 }}

 \def\refstyleprnp{              
   \gdef\refto##1{ (##1)}                                
   \gdef\refis##1{\indent\hbox to 0pt{\hss##1)~}}        
   \gdef\journal##1, ##2, ##3, 1##4##5##6{               
     {\sl ##1~}{\bf ##2~}(1##4##5##6) ##3}}

 \def\figurecaptions             
   {\endpage
    \beginparmode
    \head{Figure Captions}
 }

 \def\endpage                    
   {\vfill\eject}

 \def\endpaper                   
   {\endmode\vfill\supereject}


 \def\ref#1{Ref. #1}                     
 
 \def\frac#1#2{{\textstyle #1 \over \textstyle #2}}

 \def\sla{\raise.15ex\hbox{$/$}\kern-.57em}
 \def\leaderfill{\leaders\hbox to 1em{\hss.\hss}\hfill}
 \def\twiddle{\lower.9ex\rlap{$\kern-.1em\scriptstyle\sim$}}
 \def\bigtwiddle{\lower1.ex\rlap{$\sim$}}
 \def\gtwid{\mathrel{\raise.3ex\hbox{$>$\kern-.75em\lower1ex\hbox{$\sim$}}}}
 \def\ltwid{\mathrel{\raise.3ex\hbox{$<$\kern-.75em\lower1ex\hbox{$\sim$}}}}
 \def\square{\kern1pt\vbox{\hrule height 1.2pt\hbox{\vrule width 1.2pt\hskip
3pt
    \vbox{\vskip 6pt}\hskip 3pt\vrule width 0.6pt}\hrule height 0.6pt}\kern1pt}

 2
\def\e{{\rm e}}
\def\ltap{\raise-.55ex\hbox{\rlap{$\sim$}} \raise.4ex\hbox{$<$}}
\def\gtap{\raise-.55ex\hbox{\rlap{$\sim$}} \raise.4ex\hbox{$>$}}
\def\gsim{\mathrel{\gtap}}
\def\lsim{\mathrel{\ltap}}

\singlespace
\rightline{PUPT-1364}
\rightline{UBCTP 92-033}
\line{\hfill}
\rightline{Revised Version}
\rightline{February, 1993}
\vskip 0.5 truein
{\let\bf=\bigtenrm
\doublespace
\centerline{\bf THE SPECTRUM OF TOPOLOGICALLY MASSIVE}
\centerline{\bf QUANTUM ELECTRODYNAMICS}}
\vskip 0.5truein
\centerline{{\bf Mikhail I. Dobroliubov}$^{(a),}$\footnote{*}{Permanent
Address: Institute for Nuclear Research, Academy of Sciences of
Russia, Moscow, 117 312 Russia.}, {\bf David Eliezer$^{(b)}$},
{\bf Ian I. Kogan}$^{(c),}$\footnote{**}{Permanent Address: Institute
for Theoretical and Experimental Physics, B. Cheremushkinskaya ul.
25, Moscow, 117 259 Russia.},}
\centerline{{\bf  Gordon W. Semenoff}$^{(a)}$ and
{\bf Richard J. Szabo}$^{(a)}$}
\vskip 0.2truein
\centerline{\it $^{(a)}$Department of Physics, University of British Columbia}
\centerline{\it Vancouver, British Columbia, V6T 1Z1 Canada}
\vskip 0.2truein
\centerline{\it $^{(b)}$Department of Physics, L-412}
\centerline{\it Lawrence Livermore National Laboratory}
\centerline{\it P.O. Box 808, Livermore, California, 94551-9900 U.S.A.}
\vskip 0.2truein
\centerline{\it $^{(c)}$Joseph Henry Laboratory}
\centerline{\it Department of Physics, Princeton University}
\centerline{\it Princeton, New Jersey, 08544 U.S.A.}

\vskip 1.0truein

\centerline{\bf Abstract}
\medskip

We discuss the possibility for the spectrum of topologically massive
quantum electrodynamics with spinor matter fields to contain
unexpected and unusual stable particle excitations for certain
values of the topological photon mass. The new field
theoretical phenomena arising from this novel spectral
structure are briefly discussed.

\bigskip

{\noindent
PACS Numbers: 11.10.St, 12.20.Ds, 03.65.Ge }

\vfill\eject

\oneandahalfspace

One of the most interesting features of quantum electrodynamics in
(2 + 1) dimensions is the possibility that the photon can have a gauge
invariant topological mass term [1]. The Lagrangian contains both a Maxwell
term and a parity and time-reversal violating
Chern-Simons kinetic term for the gauge field as well as
minimal coupling to spinor matter:
$$
{\cal L}=-{1\over4e^2}F_{\mu\nu}F^{\mu\nu}+{k\over8\pi}
\epsilon^{\mu\nu\lambda}A_\mu\partial_\nu A_\lambda+\overline{\psi}
(i\gamma^\mu{\cal D}_\mu-m)\psi.\eqno(1) $$
Here $A$ is an Abelian vector field, $F_{\mu\nu}=\partial_\mu
A_\nu-\partial_\nu A_\mu$ is the field strength tensor,
${\cal D}_\mu=\partial_\mu-iA_\mu$ is the gauge covariant
derivative, and $\psi$ is a two-component Dirac field.

This model effectively
contains two dimensionful parameters, the topological photon mass
$M=e^2 k/4\pi$, which may be positive or negative depending on
the sign of the statistics parameter $k$,
and the electron mass $m$, which may also assume both positive
and negative values, but we choose $m>0$ for certainty. The topological
mass cuts off the long range Coulomb interaction which in the pure
Maxwell theory would be logarithmic.  There are two natural
dimensionless parameters in this model, the ratios $\alpha/m$ and
$\alpha/M=1/k$, where $\alpha=e^2/4\pi$. For light fermions
($m<M$) the perturbation series expansion parameter is
$\alpha^2/mM\sim1/k$, while for heavy fermions ($m\gg M$) it
turns out to be $\alpha/m=(M/m)(1/k)\ll1/k$. The standard
perturbation series can therefore be used provided $1/k\ll1$.

Drawing from our experience with QED$_4$, one would
expect that for large $k$ the spectrum of the theory (1) is
described well by perturbation theory. It contains the tree-level
particles, the photon, electron and positron, and also, because
of the attractive Coulomb interaction, metastable $e^+e^-$ bound
states, as expected from the analysis of ordinary QED$_3$ ($k\to0$ in (1)).
In the following we shall show that in fact the
qualitative properties of the spectrum are controlled
not only by the statistics parameter $k$, but also by the
parameter $\beta\equiv M/m= ke^2/4\pi m$.  When the parameter
$\beta$ is small, the
photon is relatively light, the theory resembles pure
QED$_3$, and the spectrum contains electrons, positrons,
photons and $e^+e^-$ bound states.

Also, we shall see that when $\beta$ is increased to 1,
a new phenomenon takes place -- the equal charge attraction
which was originally discussed in Ref. [2]. In this case an attractive
charge-current interaction and repulsive charge-charge and
current-current interactions become of comparable strength
and may lead to new $e^-e^-$ and $e^+e^+$ bound states [2--7].
When $M/m>1$, the $e^+e^-$ bound states no longer form.
Instead, for small enough $k\lsim1$ (the strong coupling regime),
as well as the electron, positron and photon, the spectrum
of the theory may contain scalar (if there are several
fermion flavours in the theory)
and vector $e^-e^-$ and $e^+e^+$ bound states.
We speculate
that for a special set of parameters, the vector $e^-e^-$ and
$e^+e^+$ bound states and the photon have equal masses and
form an SU(2) isospin triplet state, although at this point we have not yet
determined whether the dynamics could exhibit a full global SU(2)
symmetry.

The situation again changes at the
two-fermion threshold $M/m=2$ where the photon
becomes unstable to decay into $e^+e^-$ pairs. Here we shall show that an
unusual attraction between the electron or positron and
photon as well as $e^+e^-$ repulsion take place. There, we
conjecture that, for a certain range of parameters, the
stable particles are an $e^-\gamma$ bound state and an
$e^+\gamma$ bound state, as well as an electron and
positron and possible $e^-e^-$ and $e^+e^+$ bound states
(and then if the masses of the bound states are light enough, the
electron and positron could actually be unstable and decay into
an $e^-\gamma$ or $e^+\gamma$ bound state and a photon).
As the parameter $M/m$ becomes infinitely large, however, these unusual
particle interactions only represent fractional spin-statistics
transmutations of the composite particle systems, and the
only stable particles which remain are the electron and positron.

The $e^-e^-$ interaction in the theory (1) can be described in perturbation
theory by considering the fermion-fermion scattering amplitude in the
non-relativistic limit. Working in the tree-approximation, for different
flavour fermions we symmetrize the two corresponding $t$- and $u$-channel
diagrams to get the total symmetric amplitude $A_{\rm s}$, while for identical
fermions the Pauli exclusion principle forces us to antisymmetrize and get the
total antisymmetric amplitude $A_{\rm as}$: $$\eqalign{A_{\rm
s,as}=&-{i\over2}\Bigl[\bigl(\overline{u}(p_1')
\gamma^\mu u(p_1)\bigr)\bigl(\overline{u}(p_2')\gamma^\nu
u(p_2)\bigr)G_{\mu\nu}(p_1-p_1')\cr&\qquad\quad\pm\bigl(
\overline{u}(p_2')\gamma^\mu u(p_1)\bigr)\bigl(\overline{u}
(p_1')\gamma^\nu u(p_2)\bigr)G_{\mu\nu}(p_1-p_2')\Bigr].
\cr}\eqno(2)$$
Here $p_i$ ($p_i'$) are the momenta of the incoming (outgoing)
particles, and
$$u(p)={1\over\sqrt{2m(E-m)}}\pmatrix{E+m\cr -i(p_1+ip_2)\cr},$$
where $p=(E,p_1,p_2)$, $E^2={\vec p}\,^2+m^2$, and
$\overline{u}(p)u(p)=1$, are the on-shell
positive energy Dirac spinors [6]. We also use the (2
+ 1) dimensional representation of the Dirac matrices in terms of
Pauli spin matrices as
$\gamma_0=\sigma^3$, $\gamma_1=i\sigma^1$, and $\gamma_2=i\sigma^2$,
and
$$G_{\mu\nu}(p)=-ie^2\biggl[{p^2g_{\mu\nu}-p_\mu p_\nu\over
p^2(p^2-M^2)}
+{iM\epsilon_{\mu\nu\lambda}p^\lambda\over p^2
(p^2-M^2)}\biggr]$$
is the free photon propagator in the
Landau gauge [1].

The imaginary parts of the
$S$-matrix elements (2)
describe the lowest order long-ranged Aharonov-Bohm
flux tube interaction of the electrons,
which vanishes in the short-ranged regime and in the long
distance limit just shifts the canonical angular momentum of
the charged particles as $\ell\to\ell-{1\over k}$.
The real parts of these amplitudes, which describe the short-ranged
interactions between the electrons, in the center of mass
frame and in the non-relativistic limit are
$${\rm Re}(A_{\rm s,as})={2\pi M\over k}\biggl(1-{M\over m}\biggr)
\biggl[{1\over(\vec{p}_1-\vec{p}\,'_1)^2+M^2}\pm{1\over(\vec{p}_1
-\vec{p}\,'_2)^2+M^2}\biggr].\eqno(3)$$

We see that the
symmetrized amplitude gives a pure $S$-wave interaction ${\rm
Re}(A_{\rm
s})=4\pi(1-M/m)/kM$, while the antisymmetrized contribution is a pure
$P$-wave interaction ${\rm Re}(A_{\rm as})=4\pi
\vec{p}_1^{\,2}\cos\theta(1-M/m)/kM^3$, where $\theta$ is the
scattering angle. For any $k<0$, or $k>0$ and $M/m<1$, these amplitudes
just represent the expected repulsion
between the electrons. However, for $M/m>1$, the
magnetic attraction parts (second terms) of these amplitudes,
which arise from the Pauli dipole interaction of the electrons
due to their magnetic moment [4,6],
are stronger than the equal charge-charge
Coulomb repulsions (first terms), and
lead to a short-ranged equal charge attraction between the electrons.
This remarkable equal charge attraction implies
that for $M/m>1$ an unusual spectrum of
$S$-wave and $P$-wave $e^-e^-$ bound states may exist
in the spectrum of the quantum field theory (1),
 and it may even lead to
a possible vacuum instability in the relativistic theory [7], whereby
the vacuum could become unstable to the production of
$e^+e^-$ pairs followed by a separation of phases. It also
implies that there may be a fermion chiral condensate,
$<\overline{\psi}\psi> { } \neq0$, and the corresponding BCS
gap equation for the superconducting ground state (of electron
Cooper pairs) can be readily solved in the weak binding
case [2].

One can now obtain the potential for the $e^-e^-$ interaction
in configuration space.
We consider only the $t$-channel amplitude in (3) (i.e.
assume the fermions are distinguishable), include the imaginary
Aharanov-Bohm part
$${\rm Im}(A_{\rm s,as})={4\pi M^2\over km}
{\vec{q}\times\vec{p}_1\over\vec{q}\,^2(\vec{q}\,^2
+M^2)}$$
of (2), and write (2) as a
momentum space first order Born amplitude in the momentum transfer
$\vec{q}=
\vec{p}_1-\vec{p}\,'_1$. Notice, however, that since the
non-relativistic Hamiltonian is quadratic in momentum, the
single-photon-exchange approximation is not gauge invariant
and one must take into account the two-photon exchange
contribution. The two dimensional Fourier transform
of this resulting Born amplitude gives the
configuration space potential for the $e^-e^-$ interaction, and
then, including the centrifugal barrier, the effective potential
appearing in the corresponding radial Schr\"odinger equation is
$$V_{ff}^{(\ell)}(r)={2M\over
k}\biggl(1-{M\over m}\biggr)K_0(Mr) +{\ell_R(r)^2\over mr^2}.\eqno(4)$$

In (4), $\ell_R(r)=\ell-(1/k)\bigl[1-MrK_1(Mr)\bigr]$ is the
radially dependent
Aharanov-Bohm renormalization of the usual integer-valued
angular momentum quantum
number $\ell$ of the $e^-e^-$ pair, $K_n(x)$ denotes the
irregular modified Bessel function of order $n$ [8], and we
assume that $M>m$ in (4). The modified Bessel functions have
the asymptotic behaviours $K_n(x)\to0$ for $x\to\infty$
and $K_0(x)\sim-\log x$, $K_1(x)\sim1/x$ for
$x\to0$. Notice that the term of order
$1/k^2$ in the potential (4), which arises from the
two-photon exchange diagrams [6], was not taken into account
in Refs. [5,9], which lead the authors of those papers to
the incorrect conclusion that the effective centrifugal
barrier could be attractive. They therefore find $e^-e^-$
bound states for $\beta\ll1$, contradicting the corresponding
QED$_3$ result, which should be correspondent in this regime.
Moreover, the bound states found in Refs. [5,9] existed for
small $k$, where perturbation theory breaks down and higher
order contributions to the scattering amplitude are equally
important (so that one cannot neglect terms of order $1/k^2$
as compared to terms of order $1/k$)\footnote{$^1$}{\tenpoint
We would also like to stress that the original derivation
of the equal charge attraction which was done in Ref. [2] was both
quantum mechanical {\tenit and} field theoretical (and the
field theoretical derivation was repeated in Refs. [5,9]),
contrary to the claim that the field theoretical derivation
suggested in those papers is completely different from the
results of Ref. [2].}.

We employ a standard semi-classical analysis to
the potential (4) [10], which yields exact results for
the harmonic oscillator and hydrogen atom
Hamiltonians. In the WKB approximation,
the Bohr-Sommerfeld quantization condition for the radial
Schr\"odinger action integral reads
$$\int_{r_1}^{r_2}dr\,\sqrt{m\bigl(E_{n,\ell}-V_{ff}^{(\ell)}(r)
\bigr)}=\Bigl(n+{1\over 2}\Bigr)\pi\qquad;\qquad n=0,\,1,\,2,
\ldots,\eqno(5)$$
where $r_1$ and $r_2$ are the classical turning points defined by the
zeroes of the function $r^2\bigl(E_{n,\ell}-V_{ff}^{(\ell)}(r)\bigr)$,
$r\geq0$, and $E_{n,\ell}$ is the mass of the bound state with the
quantum numbers $n,\,\ell$.  Introducing the dimensionless parameters
$x=Mr$ and $\varepsilon_{n,\ell}=E_{n,\ell}/M$, from
(4) the quantization law (5) can be written as
$$\beta^{-1/2}\int_{x_1}^{x_2}dx\,\biggl[\varepsilon_{n,\ell}
-{2\over k}\Bigl(1-\beta\Bigr)K_0(x)-{\beta\over
x^2}\biggl(\ell-{1\over k}\Bigl(1-xK_1(x)\Bigr)\biggl)^2
\biggr]^{1/2}=\Bigl(n+{1\over2}\Bigr)\pi.\eqno(6)$$

We have solved the equation (6) numerically for $\ell=0,\,1$ and
$n=0,\,1$ using the standard method of successive approximations.  We
restricted the parameter $\beta$ to the region $1<\beta\leq15$, and
it was found that bound states (i.e. solutions $\varepsilon
_{n,\ell}$ to equation (6)) existed only for $0<k\leq1.6$. In fact,
only in the case of the lowest lying $S$-wave state ($n=\ell=0$) did
$k$ cover the full range 0--1.6 (in the other cases bound states
ceased to exist typically at about $k=0.6$). The bound state energies
$\varepsilon_{n,\ell}$ are all large for large $\beta$ and
small $k$, and as one moves away from this region the values
tail off smoothly. Notice, however, that the shift in the centrifugal
barrier introduces a large hump in the potential, allowing for
metastable bound states. In fact, all of the excited $e^-e^-$
states ($n\geq1$) were found to be metastable, and moreover
existent for only a small range of $k$-values, this parameter range
diminishing for smaller values of $\beta$ (this also occured
for $n=0$ at $\beta\sim1$ where, upon examining (4), bound states
are not expected). These numerical results also suggest a parameter
dependence of the bound state energies as $\varepsilon_{n,\ell}
\sim\beta\,\e^{-k}$.

It is easy to show that in the perturbative regime
($1/k\ll1$) and in the non-relativistic approximation
($r\gg1/m>1/M$) the potential (4) is purely repulsive and no
bound $e^-e^-$ states can form. For $1/k\gsim1$ the
perturbative calculations are not reliable, and we do not
have an expression for the potential of the interaction
between two fermions in this case. But one can see that
if one considers the perturbative potential (4) for
smaller and smaller $k$, it becomes more and more
attractive. Our numerical studies above of this potential
indicate that it can possess $S$- and $P$-wave bound
states for $k\lsim1$.

Notice also
that as $M$ becomes very large (the so-called ``anyon limit")
the short-ranged potential in (4) becomes an attractive
delta-function potential $-(4\pi/k)(1/m)\delta^{(2)}(\vec{r}\,)$,
and the barrier shift is exactly $\ell\to\ell-{1\over k}$. Although
this delta-function potential admits an $S$-wave bound
state\footnote{$^2$}{\tenpoint For a
discussion of the delta-function potential in two spatial
dimensions, see Ref. [11].}, there are no bound states in
this limit because the renormalized centrifugal barrier
is never absent.
This potential
just represents the Pauli interaction of the fermionic magnetic
moment with the flux tubes attached to the charged particles
which transmute them into anyons, and its only effect is to
give the electrons
fractional spin and statistics [4,6,12]. So, we may conclude
that as long as we can trust perturbation theory, $e^-e^-$
bound states do not exist in the theory (1). We conjecture
that they will form for $M>m$, provided $k$ is sufficiently
small, $k\lsim1$, and the photon is not too heavy\footnote{$^3$}
{\tenpoint Actually, it was shown in Refs. [4,6] that
one-loop radiative corrections give the region of dominant
magnetic attraction between two equally charged fermions
as $k<-{7\over3}$. We neglect these finite renormalizations
of all bare parameters appearing in (1).}.

In a similar fashion one can find the potential for the
$e^+e^-$ interaction. The
tree-level Feynman amplitude in the perturbation
expansion for $e^+e^-$ scattering is
given by the sum of the corresponding $t$- and $s$-channel
(annihilation) diagrams:
$$\eqalign{V_{f\bar{f}}=&\,i\Bigl[\bigl(\overline{u}(p_1')\gamma^\mu
u(p_1)\bigr)\bigl(\overline{v}(p_2)\gamma^\nu v(p_2')\bigr)
G_{\mu\nu}(p_1-p_1')\cr&\qquad\quad-
\bigl(\overline{v}(p_2)\gamma^\mu
u(p_1)\bigr)\bigl(\overline{u}(p_1')\gamma^\nu
v(p_2')\bigr)G_{\mu\nu}(p_1+p_2)\Bigr]\cr}\eqno(8)$$ where
$$v(p)={1\over\sqrt{2m(E+m)}}\pmatrix{-i(p_1-ip_2)\cr
E+m\cr}$$
are
the on-shell negative energy Dirac spinors with $\overline{v}(p)
v(p)=1$ [6].
Again the imaginary part of (8) represents the
Aharanov-Bohm flux tube interaction between the charged particles, and
in the center of mass frame and in the
non-relativistic limit, its real part reads
$${\rm Re}\bigl(V_{f\bar{f}}(\vec{q}\,)\bigr)=-{4\pi
M\over k}\biggl[\biggl( 1-{M\over
m}\biggr){1\over\vec{q}\,^2+M^2}+{1\over m(2m-M)}
\biggr].\eqno(9)$$

We see that for any $k<0$, or $k>0$ and $M/m<1$, the amplitude
(9) represents an $e^+e^-$ attraction, and the expected
$S$- and $P$-wave $e^+e^-$ bound states may appear.  For
$1<M/m<2$, the $\vec{q}$-independent ($s$-channel) term in (9) is
still attractive and dominates over the other $t$-channel term. This
term corresponds to a delta-function potential in configuration space,
which is known to have $S$-wave bound states even for weak coupling [11].
Therefore
$e^+e^-$ bound states can exist in either $S$-wave or
$P$-wave, provided that either $k<0$ or the photon is
stable.
However, if the photon is unstable, then
the $s$-channel and magnetic interaction terms in (9) dominate the
Coulomb charge attraction between the electron and positron. Then
the $e^+e^-$ pair no longer attract, and a stable
$e^+e^-$ bound state is not possible, as one would expect for unstable
photons.

As a function of the topological mass $M$, the attraction represented
by (9) is strongest just below the two-fermion threshold
$M\sim2m+0^-$, and near zero photon mass $M\sim0^\pm$
where the topologically massive gauge theory (1) degenerates into ordinary
QED$_3$, precisely where one would expect $e^+e^-$ bound states to exist.
The strongest repulsion between the electron and positron occurs
just above the two-fermion threshold $M\sim2m+0^+$, where the photon
becomes unstable and the structure of the
quantum field theory (1) begins to deviate
enormously from that of usual QED$_3$.
It is therefore expected that the spectrum of stable $e^+e^-$
bound states will be concentrated in the regions
$M\sim2m+0^-$ and $M\sim0^\pm$.
Notice also that in the
anyon limit the short-ranged interaction (9) becomes the (now
repulsive) delta-function interaction discussed before, and all that
remains of the $e^+e^-$ interaction is the pair's anomalous
statistics and the usual fractional
shift in the canonical angular momentum.

The effective configuration space interaction potential
appearing in the radial Schr\"odinger equation for the
$e^+e^-$ system corresponding to the Born amplitude (8) is
$$V_{f\bar{f}}^{(\ell)}(\vec{r}\,)=-{2M\over k}\biggl[\biggl(1-{M\over
m}\biggr) K_0(Mr)+{2\pi\over
m(2m-M)}\delta^{(2)}(\vec{r}\,)\biggr]
+{\ell_R(r)^2\over mr^2}\eqno(10)$$
where we take
$0<M/m<2$ in (10). Notice that in this regime and for
$1/k\ll1$ the barrier shift is small and it vanishes for
$M\to0$. Now the delta-function in (10) acts only
on $S$-wave states as a constant shift in the corresponding $S$-wave
energy eigenvalues. Therefore this term, which represents the $s$-channel
annihilation of the electron and positron, can be neglected in the
ensuing bound state analysis, remembering to just shift all $S$-wave
energy levels by $-(M^2/m)\bigl(\e^{k(2m-M)/2M} -1\bigr)^{-1}$. We are
therefore left with the same Bohr-Sommerfeld equation (6) to solve,
except that now $\beta<1$ and
 of course the sign of the Bessel function $K_0(x)$ in
(6) changes. A similar numerical analysis now as that of (6)
supports the idea of the existence of $S$- and $P$-wave $e^+e^-$
bound states in this regime of the theory.

A somewhat more analytic analysis of equation (6) for the
$e^+e^-$ system can be performed near the zero photon mass
threshold $M\to0$ (which also gives the corresponding result in
ordinary QED$_3$ where $M$ then plays the role of an infrared cutoff).
For $M\ll m$, the short-ranged
potential in (10) is nothing but the attractive logarithmic Coulomb
potential in two spatial dimensions, which is known to have a discrete
eigenvalue spectrum bounded from below with eigenvalues of finite
multiplicity [13]. In this case the equation (6) can be written
to leading order in $\beta$ as
$$\int_{y_1}^{y_2}dy\,\biggl[-\ell^2-{2\over k\beta}
\,\e^{k\varepsilon_{n,\ell}}\Bigl(
1-\beta\Bigr)y\,\e^{2y}\biggr]^{1/2}+{\rm O}(\beta^{1/2})
=\Bigl(n+{1\over2}\Bigr)\pi,$$
where $y=\log x-k\varepsilon_{n,\ell}/2(1-\beta)$ and $y_1$
and $y_2$ are the two zeroes of the function $\ell^2
+(2/k\beta)(1-\beta)\,\e^{k\varepsilon_{n,\ell}}
y\,\e^{2y}$, $y\in{\bf R}^1$, so that $y_1\sim-\infty$
and $y_2\sim0$ for $\beta\to0$. This expression
can be integrated with relative ease and evaluated for
$\ell=0,\,1$
to yield approximate
expressions for the masses of the $S$- and $P$-wave $e^+e^-$ bound states:
$$\eqalign{E_{n,\ell=0}^{(f\bar{f})}&\simeq{2M\over k}
\log\biggl[\sqrt{\pi kM\over m}\Bigl(n+{1\over2}\Bigr)\biggr]
\cr E_{n,\ell=1}^{(f\bar{f})}&\simeq{2M\over k}\log\biggl[
\biggl({3\sqrt{2\e}\over\sqrt{\e}+3\sqrt{2}}\biggr)
\sqrt{Mk\over2m}\Bigl(n+{1\over2}\Bigr)\pi\biggr].\cr}\eqno(11)$$
Then (11) also shows that the number of $S$- and $P$-wave $e^+e^-$
bound states are $N_{\ell=0}\sim\sqrt{m/\pi kM}$ and
$N_{\ell=1}\sim\sqrt{2m/kM}$ respectively, which are both quite large.

Thus for $M/m<2$ the $e^+e^-$
spectrum of states is qualitatively similar to that expected in ordinary
QED$_3$, with quite a large number of $e^+e^-$ bound states appearing.
Although the existence of $e^+e^-$ bound states in topologically massive
quantum electrodynamics for stable photons may not seem perplexing,
the above results being standard in the analysis of the two dimensional
Coulomb potential, the existence of another stable and neutral vector
particle other than the stable photon can lead to
interesting consequences for the structure of the quantum field theory (1).
In particular, although it is known from the Coleman-Hill theorem [14] that,
when the matter fields have a finite mass gap [15],
the statistics parameter does not renormalize beyond one-loop order in
perturbation theory with respect to the topologically
massive {\it photon}, it could be
further renormalized by the $P$-wave $e^+e^-$ bound state (for
one knows that vector particles in general do contribute to
this renormalization [16]).

The final elementary interaction one could consider for the
quantum field theory (1) is the Compton scattering of an electron and a
photon. The photon, being a massive vector excitation in this theory,
allows us to treat this two-body interaction using standard methods,
as above. The total Compton $S$-matrix element in the tree
approximation is the sum of the $s$- and $u$-channel diagrams for the
$e^-\gamma$ interaction:
$$\eqalign{V_{f\gamma}=&-i\Bigl[\bigl(\overline{u}(p_1')
\gamma^\nu S(p_1+p_2)\gamma^\mu u(p_1)\bigr)e_\mu(p_2)
e_\nu^*(p_2')\cr&\qquad\quad+\bigl(\overline{u}(p_1')
\gamma^\nu S(p_1-p_2')\gamma^\mu u(p_1)\bigr)e_\nu(p_2)
e_\mu^*(p_2')\Bigr],\cr}\eqno(12)$$ where
$S(p)=\,i(p_\mu\gamma^\mu-m)^{-1}$ is the free fermion propagator, and
$$e(p)={1\over\sqrt{2}M\vert\vec{p}\,\vert}\pmatrix{\vec{p}\,^2\cr
Ep_1-iMp_2\cr Ep_2+iMp_1\cr}$$
are the on-shell polarization vectors for the photons in the
transverse Landau gauge with $p\cdot e(p)=0=e(p)\cdot e(p)$
and $e(p)\cdot e^*(p)=-1$ [17].
In the center of mass frame, it is found
that only the $u$-channel (second term) part of (12) contributes in
the non-relativistic limit:
$$V_{f\gamma}(\vec{q}\,)=-{2M\over\vec{q}\,^2+\mu^2}\cos\theta
+{4iM\,{\rm sign}(M)\over\vec{q}\,^2(\vec{q}\,^2+\mu^2)}
\vec{q}\times\vec{p}_1(1-\cos\theta),\eqno(13)$$
where we have introduced the mass parameter $\mu=
\sqrt{M(M-2m)}$.

In the extreme non-relativistic limit $\vec{q}\,^2\to0$ this
potential becomes
$$V_{f\gamma}\Bigm\vert_{\vec{q}\,^2\to0}={2\over2m-M}
\,\e^{i\,{\rm sign}(M)\theta}.\eqno(14)$$
Thus, for stable photons the $e^-\gamma$ pair repel each other, no
bound state can form, and the entire spectrum of the quantum field
theory (1) is identical in nature to that in ordinary (2 + 1) (or (3 + 1))
dimensional quantum electrodynamics. But if the photon is unstable,
the $e^-\gamma$ pair attract each other in $P$-wave, and an
$e^-\gamma$ bound state may form. It is in this unstable region
$M/m>2$ that the spectrum of the theory (1) becomes somewhat exotic and
differs enormously from what one would usually expect.

The interaction amplitude (14) describes an extremely strong
attraction between the electron and photon for values of the
topological mass just above the two fermion threshold, $M\sim2m+0^+$.
Just below this threshold, the electron and photon repel each other
very strongly (notice that this is exactly opposite to the case of the
$e^+e^-$ interaction and is similar to the $e^-e^-$
interaction).  For a binding energy $E=-E_b^{(f\gamma)}<0$, it is easy
to see that the condition for the formation of a stable $e^-\gamma$
bound state is just $E_b^{(f\gamma)}>M-2m$, and thus stable
$e^-\gamma$ bound states could form even though the photon itself
is not a stable particle in this regime of the
theory.  Also notice that in the anyon limit
the interaction amplitude (14) vanishes,
and the stability condition $E_b^{(f\gamma)}>M-2m$ is violated.
Therefore no stable $e^-\gamma$ bound state can form in the anyon
limit.

The Fourier transform of (13) gives the non-central
configuration space effective interaction potential
$$\eqalign{V_{f\gamma}^{(\ell)}(r,\phi)=&-{2M\over\pi\mu^2r^2}
\biggl\lbrace\ell\bigl(1-\mu rK_1(\mu r)\bigr)
+\cos2\phi\biggl[\biggl({2\ell\over\mu r}(\mu^2r^2-2) -\mu
r\biggr)K_1(\mu r)\cr&\qquad\qquad-(\mu^2r^2+2\ell) K_0(\mu
r)+1\biggr]\biggr\rbrace+{\ell^2\over 2m^*r^2}\cr}\eqno(15)$$
in the $e^-\gamma$ radial Schr\"odinger
equation, where $(r,\phi)$ are the polar
coordinates of the relative $e^-\gamma$ position
vector $\vec{r}$ and $m^*=mM/(m+M)$ is the reduced mass of
the $e^-\gamma$ system. We assume that $M>2m$ in (15).

Notice that the imaginary part of (13) is similar to (but is not
exactly) the Aharanov-Bohm interaction amplitude.  Notice also that in
the anyon limit $M\gg2m$, $\mu\to\infty$, and the expression (15)
becomes
$V_{f\gamma}^{(\ell)}(r,\phi)\Bigm
\vert_{\mu\to\infty}=-(2M/\pi\mu^2r^2)(\ell+\cos2\phi)
+\ell^2/2mr^2$.
This represents
a shift in the orbital angular
momentum of the $e^-\gamma$ composite system of
$\ell\to\ell- {2m\over\pi M}$. As $M\gg2m$ this shift is
small and negligible except for the $S$-wave channel ($\ell=0$),
just as for the $e^-e^-$ system when $k$ is large.
Therefore the $e^-\gamma$ system has an unusual spin-statistics
transmutation in the anyon limit
which is small, but
nevertheless a non-trivial adiabatical rotation through $2\pi$ of the
$e^-\gamma$ system as a function its radius $r$ exists and can be
computed via (14).

The other limiting case is $M\sim2m+0^+$ ($\mu\to0^+$). In this
limit
$$V^{(\ell)}_{f\gamma}(r,\phi)\Bigm
\vert_{\mu\to0^+}=-{2M\over\pi\mu^2r^2}\cos2\phi
\biggl[1+2\ell\biggl(\log\mu r-{2\over(\mu r)^2}\biggr)\biggr]
+{3\ell^2\over4mr^2}.\eqno(16)$$
This potential (without the centrifugal barrier) is completely
angle dependent, and it is here that the most and
strongest $e^-\gamma$ bound states are expected to be concentrated.

In the general case (15), as well as in the limiting case (16),
an analysis of the corresponding
Schr\"odinger equation as for the $e^-e^-$ and
$e^+e^-$ systems appears to be quite difficult due to the
non-trivial angular dependence of $V_{f\gamma}^{(\ell)}
(r,\phi)$, which has the form of a dipole or quadrupole interaction.
Moreover, it is not clear how this angular dependence
affects the binding energies. Nonetheless,
a contour plot shows that the potential (15)
has very deep wells in certain regions of the polar
$(r,\phi)$ plane, and it is thus not unreasonable to speculate
that at least for a certain set of parameters, the potential
(15) has stable bound states.
It is also interesting, from examining the small $r$ divergent behaviour of
the Bessel functions in (15), that the $e^-\gamma$ system can
undergo a falling into the center. Moreover, if $e^-\gamma$
and $e^+\gamma$ bound
states exist, then the relativistic vacuum attracts electrons
and positrons to
itself, and an ensuing vacuum instability can occur, much in the same
way as for the unusual $e^-e^-$ attraction. Whether or not the
charged composite $e^-\gamma$ bound state can keep on attracting
electrons or positrons to itself is another question.  This provides
evidence that the analysis of the quantum field
theory (1) {\it must} be regarded as a true many-body quantum field theory
problem (not just a quantum mechanical one), and the possibility of
$e^-\gamma$  and $e^+
\gamma$ bound states in this theory represents a new novel
field theoretical phenomenon.

We see therefore that the spectrum of topologically massive
spinor electrodynamics separates into different phases, with
critical phase transition points at $M/m=1$ and $M/m=2$
separating the expected spectrum from the exotic one. In the
limit of a small topological mass, the spectrum contains only
the usual particles expected from ordinary QED$_3$ (or QED$_4$),
while at these critical phase transition points unusual
electron and photon composite particles can emerge in the
spectrum. In the anyon limit, the only effect of these particle interactions
is to give the composite particle systems
exotic spin-statistics transmutations,
as expected from the effective
quantum field theories of anyonic systems [12], and the
only stable particle excitations which survive are those
induced by the fundamental fermion field $\psi$.

We close by noting that in the scalar version of the quantum
field theory (1), the exotic spectral structure discussed
above is not possible. Consider the Lagrangian (1), except
now with gauge coupling to scalar matter fields $\varphi$:
$${\cal L}_S=-{1\over4e^2}F_{\mu\nu}F^{\mu\nu}+{k\over8\pi}
\epsilon^{\mu\nu\lambda}A_\mu\partial_\nu A_\lambda+
({\cal D}_\mu\varphi)^{\dagger}({\cal D}^\mu\varphi)-m_S^2
\varphi^\dagger\varphi,\eqno(17)$$
where $m_S$ is the meson mass. The meson-meson,
meson-antimeson, and meson-photon $S$-matrix elements in the
tree-approximation are respectively
$$\eqalign{V_{mm}=&\,{\rm Re}\bigl[-i(p_1+p_1')^\mu(p_2+p_2')^\nu
G_{\mu\nu}(p_1-p_1')\bigr]={16\pi M\over k}{m_S^2\over
\vec{q}\,^2+M^2}\cr
V_{m\bar{m}}=&\,{\rm Re}\bigl[i(p_1+p_1')^\mu
(p_2+p_2')^\nu G_{\mu\nu}(p_1-p_1')
+i(p_1-p_2)^\mu(p_1'-p_2')^\nu
G_{\mu\nu}(p_1+p_2)\bigr]\cr=&-{16\pi M\over k}\biggl(
{m_S^2\over\vec{q}\,^2+M^2}+{\vec{p}_1\cdot\vec{p}\,'_1\over
M^2-4m_S^2}\biggr)\cr
V_{m\gamma}=&-i\bigl[(2p_1+p_2)^\mu
(2p_1'+p_2')^\nu e_\mu(p_2)e_\nu^*(p_2')D(p_1+p_2)\cr&\qquad+
(2p_1-p_2')^\mu(2p_1'-p_2)^\nu e_\nu(p_2)e_\mu^*(p_2')
D(p_1-p_2')-2ig^{\mu\nu}e_\mu(p_2)e_\nu^*(p_2')\bigr]\cr
=&2\,\e^{i\,{\rm sign}(M)\theta}\cr}\eqno(18)$$
in the center of mass frame and in the non-relativistic limit.
Here $D(p)=i(p^2-m_S^2+i\epsilon)^{-1}$ is the free scalar
propagator, and we have taken $\vec{q}\,^2\to0$ in the final
amplitude $V_{m\gamma}$ of (18).

The meson-meson pair
therefore always repel for any values of the parameters, while
the meson and antimeson always attract each other. This occurs
because spinless particles have no magnetic moment at tree-level,
and hence no unusual Chern-Simons magnetic interaction
between them, and they therefore only interact through the
usual Coulombic forces (unlike spinor particles). The meson-photon
pair also always repel each other in $P$-wave, since in the
relevant limits the repulsive bare four-point vertex (third term
of $V_{m\gamma}$ in (18)) dominates the total meson-photon interaction
(whereas in the spinor case the photon can only dress
the fermions).
Thus topologically massive scalar electrodynamics can not admit
exotic bound states, and the spectrum of the scalar quantum
field theory (17) contains precisely the same particles as the
ordinary three (or four) dimensional scalar Maxwell theory does.

\bigskip

{\noindent{\it Acknowledgements: } The authors wish to thank M. Bergeron and A.
Vainshtein for helpful discussions. The work of M. D., G. S. and R. S. was
supported in part by the Natural Sciences and Engineering Research Council of
Canada.  The work of I. K. was supported by the U. S. National Science
Foundation grant \# NSF PHY90-21984 and that of D. E. was supported
by a grant from the U. S. Department of Energy.}

\vfill\eject

\singlespace
\centerline{\bf References}
\bigskip

\item{[1]} J. F. Schonfeld, Nucl. Phys. {\bf B185} (1981), 157;
S. Deser, R. Jackiw and S. Templeton, Phys. Rev. Lett. {\bf 48}
(1982), 475; Ann. Phys. (N.Y.) {\bf 140} (1982), 372.
\line{\hfill}
\item{[2]} I. I. Kogan, JETP Lett. {\bf 49} (1989), 225.
\line{\hfill}
\item{[3]} A. Groshev and E. R. Poppitz, Phys. Lett. {\bf B235}
(1990), 336.
\line{\hfill}
\item{[4]} I. I. Kogan, Phys. Lett. {\bf B262} (1991),
83; J. Stern, Phys. Lett. {\bf B265} (1991), 119; I. I. Kogan
and G. W. Semenoff, Nucl. Phys. {\bf B368} (1992), 718.
\line{\hfill}
\item{[5]} H. O. Girotti, M. Gomes and A. J. da Silva,
Phys. Lett. {\bf B274} (1992), 357.
\line{\hfill}
\item{[6]} R. J. Szabo, I. I. Kogan and G. W. Semenoff,
Nucl. Phys. {\bf B} [FS] (1992), in press.
\line{\hfill}
\item{[7]} I. I. Kogan and I. V. Polyubin, SSC preprint
SSCL-SR-284 (1990) (unpublished).
\line{\hfill}
\item{[8]} I. S. Gradshteyn and I. M. Rhyzhik, {\it Table of
Integrals, Series, and Products}, Academic Press (San Diego)
(1980).
\line{\hfill}
\item{[9]} H. O. Girotti, M. Gomes, J. L. deLyra, J. R. S.
Nascimento and A. J. da Silva, Phys. Rev. Lett. {\bf 69} (1992),
2623; ``Electron-Electron Bound States in QED$_3$", University
of Sao Paulo preprint IFUSP/P-1012 (1992).
\line{\hfill}
\item{[10]} A. Messiah, {\it Quantum Mechanics I}, John
Wiley and Sons (New York) (1976).
\line{\hfill}
\item{[11]}
K. Huang, {\it Quarks, Leptons,
and Gauge Fields}, World Scientific (Singapore) (1982).
\line{\hfill}
\item{[12]} F. Wilczek and A. Zee, Phys. Rev. Lett. {\bf 51}
(1983), 2250; G. W. Semenoff, Phys. Rev. Lett. {\bf 61} (1988),
517; T. Matsuyama, Phys. Lett. {\bf B228} (1989), 99; G. W. Semenoff
and P. Sodano, Nucl. Phys. {\bf B328} (1989), 753; S. Forte
and T. Jolicoeur, Nucl. Phys. {\bf B350} (1991), 589.
\line{\hfill}
\item{[13]} F. Geszetesy and L. Pittner, J. Phys. {\bf A11}
(1978), 679.
\line{\hfill}
\item{[14]} S. Coleman and B. Hill, Phys. Lett. {\bf B159}
(1985), 184; Y. Kao and M. Suzuki, Phys. Rev. {\bf D31} (1985),
2137; M. Bernstein and T. Lee, Phys. Rev. {\bf D32}, (1985), 1020.
\line{\hfill}
\item{[15]} G. W. Semenoff, P. Sodano and Y.-S. Wu,
Phys. Rev. Lett. {\bf 62} (1989), 715.
\line{\hfill}
\item{[16]} R. Jackiw and S. Templeton, Phys. Rev. {\bf D23}
(1981), 2291.
\line{\hfill}
\item{[17]} I. I. Kogan and A. Morozov, Sov. Phys. JETP
{\bf 61} (1985), 1.

\end